\documentclass{elsart}
\usepackage{epsfig}
\usepackage{amssymb}
\usepackage{amsmath}
\usepackage{cite}

\def\eg{{\it e.g.~}}

\def\half{\frac{1}{2}}

\begin{document}

\begin{frontmatter}

% Title, Authors And Addresses

% Use The Thanksref Command Within \Title, \Author Or \Address For Footnotes;
% Use The Corauthref Command Within \Author For Corresponding Author Footnotes;
% Use The Ead Command For The Email Address,
% And The Form \Ead[Url] For The Home Page:
% \Title{Title\Thanksref{Label1}}
% \Thanks[Label1]{}
% \Author{Name\Corauthref{Cor1}\Thanksref{Label2}}
% \Ead{Email Address}
% \Ead[Url]{Home Page}
% \Thanks[Label2]{}
% \Corauth[Cor1]{}
% \Address{Address\Thanksref{Label3}}
% \Thanks[Label3]{}

%\Title{Traffic Jams And Ordering In Driven Systems}
\title{Traffic jams and ordering far from thermal equilibrium}


\author[WIS]{E. Levine},
\author[WIS]{G. Ziv},
\author[MIN]{L. Gray}, and
\author[WIS]{D. Mukamel}
\address[WIS]{ Department of Physics of Complex Systems, Weizmann
Institute of Science, Rehovot, Israel 76100.} \address[MIN]{School
of Mathematics, University of Minnesota, Minneapolis, Minnesota
55455.}

\begin{abstract}
The recently suggested correspondence between domain dynamics of
traffic models and the asymmetric chipping model is reviewed. It
is observed that in many cases traffic domains perform the two
characteristic dynamical processes of the chipping model, namely
chipping and diffusion. This correspondence indicates that jamming
in traffic models in which all dynamical rates are
non-deterministic takes place as a broad crossover phenomenon,
rather than a sharp transition. Two traffic models are studied in
detail and analyzed within this picture.
\end{abstract}

\end{frontmatter}

% main text
\section{Introduction}
\label{sec:intro}

Ordering in one-dimensional systems far from thermal equilibrium
has been studied extensively in recent years
\cite{Mukamel00,Evans00,GunterRev}. In the case of thermal
equilibrium, it is well known that one-dimensional systems with
short-range interactions cannot exhibit long range order. In
contrast, it has been repeatedly demonstrated that non-equilibrium
driven systems, whose dynamics does not obey detailed-balance, can
be ordered even when the dynamics is local and noisy.

Single-lane traffic constitutes a particularly interesting class
of one-dimensional driven systems. Modeling traffic dynamics has
been of considerable interest over the years
\cite{Chowdhury00,Helbing01}. A very useful quantity which has
often been used to characterize traffic flow is the relation
between the density of cars in the road and the traffic
throughput. This relation, termed the {\em fundamental diagram},
was measured empirically in various situations, and was studied in
a large variety of models \cite{Chowdhury00,Helbing01,Hall}. At
low car density one expects the traffic to flow smoothly with a
linear increase of the throughput with the car density. On the
other hand at high densities jams are formed and the flow is
lowered, sometimes even to a complete stop \cite{Hall,Brilon}. One
is interested in developing a better and deeper understanding of
the dynamical mechanisms which are involved in this behavior. In
particular, an intriguing question is whether jamming takes place
via a genuine sharp phase transition at a particular car density,
or perhaps it develops as a broad crossover phenomenon
\cite{Chowdhury00,Helbing01,Kertesz}. This question is closely
related to the question of ordering in driven one-dimensional
systems.

In recent years probabilistic Cellular Automata (CA) models have
been introduced  to analyze traffic flow
\cite{Chowdhury00,Helbing01,Cremer86,Nagel92}. In such models both
time and space are discrete and all dynamical variables  (\eg
position and velocity of all cars) are updated simultaneously
according to some update scheme. This provides a rather efficient
way for carrying out numerical studies of the fundamental diagram.
Some traffic CA models were suggested to exhibit jamming phase
transitions \cite{Nagel95,Krauss97,Eisenblatter98,Lubeck98,VDR}.
However, the existence of such a transition can only be explicitly
demonstrated in some limiting cases where certain dynamical
processes are deterministic. These cases are less relevant for
realistic traffic flow where all dynamical processes are expected
to be noisy. The existence of a jamming phase transition in more
generic cases where all dynamical processes are non-deterministic
is a more difficult theoretical question.

Recently it has been suggested \cite{Levine03} that coarse-grained
traffic dynamics can be modeled by the asymmetric Chipping Model
(CM)
\cite{Dennis88,Krapivsky96,Majumdar98A,Majumdar98B,Majumdar98C,Rajesh02}.
This model, introduced a few years ago, yields a particular
mechanism of condensation transition in one-dimensional models.
The CM belongs to a class of urn models. These are simple lattice
models, defined on a ring geometry, where each site can either be
vacant or occupied by one or more particles. The dynamics of the
CM involves two processes: {\em chipping}, where a single particle
hops to a nearest neighbor site at a constant rate $\omega$ ; and
{\em diffusion}, where all particles in a site hop together to an
adjacent site with rate $\alpha$. Mean-field analysis indicates
that this model exhibits a condensation transition at a critical
density. This result remains valid beyond the mean-field
approximation as long as the chipping process is symmetric, with
equal right and left hopping probabilities
\cite{Majumdar98A,Majumdar98B,Majumdar98C}. It has also been
argued that if the chipping process is biased, no condensation
takes place \cite{Rajesh02}.

%The first mechanism is described in terms of the Zero Range
%Process (ZRP) \cite{Evans00,ZRP}. In this model particles hop
%between nearest neighbor lattice sites with rates $\omega_k$ which
%depends only on the number of particles $k$ at the departure site.
%If the rates decay to zero as $k$ is increased, or if the rates
%decay slowly enough to a finite value, a condensation transition
%takes place, whereby a single lattice site becomes macroscopically
%occupied as the density is increased beyond a critical value. It
%has been suggested that the coarse-grained dynamics in a broad
%class of one-dimensional driven models can be described by a ZRP
%with rates which at large $k$ decay as $\omega_k =
%\omega_\infty\left(1+b/k\right)$ \cite{Kafri02A,Kafri03}. In this
%case phase separation can occur only if $b>2$.
%

To make the correspondence between traffic models and the CM it
has been suggested that the coarse-grained evolution of traffic
models is described in terms of domain dynamics, which essentially
involves diffusion and asymmetric chipping processes
\cite{Levine03}. It has thus been concluded that no phase
separation transition should be expected in this class of models,
in the case where all transition rates are non-deterministic. In
these cases jamming takes place as a broad crossover process
rather than via a sharp phase transition.

In this paper we review the correspondence between traffic models
and the asymmetric CM and closely examine the coarse-grained
dynamics of some traffic models in view of this correspondence. In
Section~2 we begin by introducing a simple traffic cellular
automaton, whose dynamics can be mapped onto a zero range process
(ZRP) \cite{Evans00,ZRP}. In this urn model only chipping takes
place but no diffusion. This particular traffic model is thus
restricted in its dynamics. Nevertheless, the exact solution of
this model yields an insight into the fundamental diagram which
characterizes traffic models. A more general traffic model, which
does correspond to the CM with both chipping and diffusion
processes, is introduced and studied in Section~3. To examine this
approach within a broader scope we apply it in Section~4 to a
recently introduced traffic CA \cite{Gray01}. Our results are
summarized in Section~5.

\section{A simple traffic CA -- chipping without diffusion}

We start by considering an exactly soluble traffic model, whose
domain dynamics involves only chipping processes, without
diffusion. The steady-state properties of this model are obtained
by a mapping onto a zero range process (ZRP) \cite{Evans00,ZRP}.
The model is defined on a periodic lattice of size $L$ with
$N=\rho L$ cars. It evolves in discrete time by simultaneously
updating all sites using the transition probabilities
\begin{align}\label{eq:PQ}
\bullet\bullet\circ\mathop{\rightarrow}\limits^{r}
\bullet\circ\bullet &&
\circ\bullet\circ\mathop{\rightarrow}\limits^{q}\circ\circ\bullet\,,
\end{align}
where $\bullet$ denote a car and $\circ$ a vacancy. This is a
special case of the model considered in \cite{Gray01}. It is also
closely related to the VDR model introduced and studied in
\cite{VDR}. With $r<q$ the model exhibits the 'slow-to-start'
feature which characterizes many traffic models. This simple model
belongs to a class of models where no explicit velocity variable
is attached to each car (see \eg \cite{Gray01,Antal00,Hager01}).
Rather, cars progress according to rules which depend only on the
position of neighboring cars. This class of models provides a
useful tool for studying specific features of traffic models.
Other more complex models introduce velocity variables, and define
rules by which both velocity and position are updated
\cite{Cremer86,Nagel92,VDR,Levine03}.

To proceed we define the ZRP dynamics, and argue that the dynamics
\eqref{eq:PQ} corresponds to a particular choice of its rates. The
ZRP is an urn model, where particles hop between nearest neighbor
urns with rates $\omega(k)$ which depends only on the number of
particles $k$ at the departure site. To map the traffic model onto
a ZRP it is instructive to view each vacancy as an urn, occupied
by the uninterrupted sequence of particles located at its left.
For example, the following traffic configuration corresponds to 3
urns occupied by 3, 0, and 1 particles, respectively,
$$\bullet\bullet\bullet\circ\circ\bullet\circ \qquad
\Longleftrightarrow \qquad
\begin{picture}(30,30)(0,0)
\put(0,0){$\circ$}\put(8,0){$\circ$}\put(16,0){$\circ$}
\put(0,8){$\bullet$}\put(0,16){$\bullet$}\put(0,24){$\bullet$}\put(16,8){$\bullet$}
\end{picture} \;.$$
%\begin{picture}(0,0)(25,10)
%\put(-5,-5){\line(90,0){10}}\put(0,0){\circle{6}}\put(0,10){\circle{6}}\put(0,20){\circle{6}}
%\put(8,-5){\line(90,0){10}}\put(21,-5){\line(90,0){10}}\put(24,0){\circle{6}}
%\end{picture} \;.$$
Therefore, the dynamics \eqref{eq:PQ} corresponds to hopping of
the topmost particle from an urn to its right neighbor, with rates
\begin{equation}\label{eq:PQ_mapping}
\omega(k)=\begin{cases} q & k=1\\ r & k>1\end{cases}\,.
\end{equation}

The grand-canonical steady-state distribution of the ZRP in
parallel dynamics is a product measure, whereby the single-site
occupation weight is given by~\cite{Evans00,Evans97}
\begin{equation}\label{eq_apx_zrp}
f(k)=\left\{ \begin{array}{lr} 1 &\qquad k=0 \\
\frac{z^k}{1-\omega(k)}\prod_{m=1}^k\frac{1-\omega(m)}{\omega(m)}&\qquad
k>0\end{array}\right.\,.
\end{equation}
Here $z$ serves as a fugacity, which is set to determine the
overall density. Using this solution, the un-normalized domain
size (or jam length) distribution in the traffic model with $q
\neq 1$ is
\begin{equation}\label{eq:PQ_PD}
f(k)= \begin{cases} 1 &  k = 0 \\ \frac{z}{q} & k = 1 \\
z\frac{1-q}{r\,q}\left(z\frac{1-r}{r}\right)^{k-1} & k \ge 2
\end{cases}\,.
\end{equation}
The case $q=1$ is unique, and will be treated below. The
fundamental diagram is obtained by inserting \eqref{eq:PQ_PD} in
the general expression for the current
\begin{equation}
J(z) = \frac{\sum_{k=1}^\infty \omega(k)\,f(k)}{\sum_{k=0}^\infty
(k+1)\,f(k)}\,,
\end{equation}
with the fugacity $z$ determined from the equation $\phi = \sum_k
kf(k)/\sum_kf(k)$. Here $\phi$ is the average jam-size, which is
related to the car density $\rho$ in the traffic model by
$\phi=\rho/(1-\rho)$.

\begin{figure}[tb]
\centerline{\epsfig{file=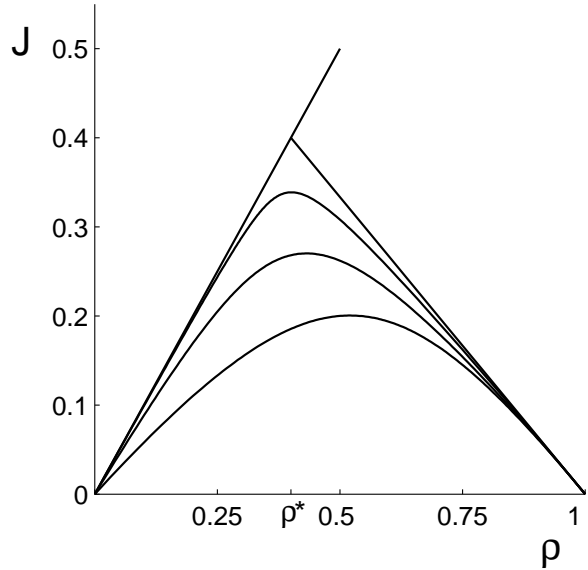,width=8truecm}}
\caption{The fundamental diagram of the simple model, for $r=2/3$
and $q=0.6, 0.9, 0.99,$ and $1$ from bottom to top. }
\label{fig:fundamental}
\end{figure}

It is readily seen from \eqref{eq:PQ_PD} that for $q \neq 1$ the
jam-length distribution decays exponentially at any density, and
thus no phase transition takes place. The fundamental diagram for
various values of $q$ at $r=2/3$ is given in
Fig.~\ref{fig:fundamental}.

\subsection{The deterministic case $q=1$}
At $q=1$ isolated particles move deterministically. This is an
example of the cruise control (CC) limit of traffic models,
whereby cars moving at their maximal speed do not decelerate as
long as they are not interrupted by other cars. We first address
the low-density regime $\rho<1/2$, where a free-flow state is
reached for any finite system at sufficiently long time. In this
state each car is separated from a neighboring car by at least one
vacancy. This is an absorbing state, since once formed it remains
unchanged. Using \eqref{eq:PQ_PD} one obtains the following
domain-size distribution
\begin{equation}
\label{eq:PQ1}
f(k)= \begin{cases} 1 &  k=0 \\ z & k = 1 \\
0 & \text {otherwise}\end{cases}\;.
\end{equation}
The car density is then given by $\rho=z/(1+2\,z)$, implying $\rho
\leq 1/2$, and a current $j(\rho)=\rho$. Clearly, at high
densities this distribution does not describe the steady state,
and one expects a jammed state, with a macroscopic jam coexisting
with a free-flow region. Since this macroscopic jam emits
particles with probability $r$, the particle density in the free
flow region is just $r/(1+r)$. It is straightforward to show that
in this case the current is given by $j(\rho)=r(1-\rho)$. The
current-density relation in the two regimes is depicted in
Fig.~\ref{fig:fundamental}, where the two branches intersect at
$\rho^*=r/(1+r)$.

In the region $\rho^*<\rho<1/2$ the two branches coexist. In this
density interval any finite system will evolve into the free-flow
branch in the long-time limit. In the following we show that in
the thermodynamic limit the time it takes for the jammed state to
reach the free-flow state grows exponentially with the system
size. This implies that in fact in these densities {\em both}
free-flow and jammed states are thermodynamically stable. To
demonstrate this point, we consider a system of length $L$
occupied by $N$ cars, with initial condition whereby all cars
reside in a single domain. With $N/L < 1/2$ the system will
eventually evolve into a free flow state. Let us estimate the time
it takes to reach this state. Cars are emitted from the front end
of the domain with probability $r$. Once a car is emitted, it
moves deterministically with velocity $1$, rejoining the domain at
its back end after exactly $L-N$ time steps (assuming that the
domain has not disintegrated in this time interval). The
probability $p_{\mbox{dis}}$ that a domain of size $N$
disintegrates during a time interval of $L-N$ is
\begin{equation}
p_{\mbox{dis}} =
\sum_{k=N-1}^{L-N}\binom{L-N}{k}r^{k}(1-r)^{L-N-k}\;.
\end{equation}
For $\rho<\rho^*$ this probability is equal to $1$ for large $L$
(as the dominant term in the sum is obtained at some $k>N$), while
in the interval $\rho^* < \rho <1/2$ the sum is dominated by the
$k=N-1$ term. Hence,
\begin{equation}
p_{\mbox{dis}}\simeq
\left[\frac{1-\rho}{1-2\rho}(1-r)\right]^{L-N}\left[\frac{\rho}{1-2\rho}\frac{1-r}{r}\right]^{-N}\;.
\end{equation}
This probability is exponentially small in $L$, and therefore the
decay time of the jammed state is exponentially long.

The fundamental diagram of this model, Fig.~\ref{fig:fundamental},
has the characteristics of a more general class of traffic models.
It exhibits a jamming transition at $\rho=\rho^*$ only in the
cruse control limit $q=1$ where at least one of the dynamical
processes become deterministic. The free-flow and the jammed
states coexist within some density interval as two
thermodynamically stable states. Once all rates become noisy, the
jamming transition disappears and the fundamental diagram becomes
smooth.

\section{Extended traffic CA -- diffusion and chipping}
\label{sec:traffic}

We now consider a more general traffic CA, whose dynamics involves
both chipping and diffusion processes. It is argued that the
domain dynamics of this model is characteristic of many traffic
models. The model is defined on a periodic lattice of size $L$,
occupied by $N=\rho L$ cars. The transition probabilities are
given by \footnote{A very similar model can be defined as a
variant of the VDR model \cite{VDR} in terms of velocities and
deceleration probabilities. For simplicity we adopt the notation
given above. }
\begin{equation}
\label{eq:smrates} \bullet\circ\circ \mathop{\longrightarrow}^a
\circ\bullet\circ \qquad \bullet\circ\circ
\mathop{\longrightarrow}^u \circ\circ\bullet \qquad
\bullet\circ\bullet \mathop{\longrightarrow}^s
\circ\bullet\bullet\;,
\end{equation}
where as before, $\bullet$ denotes a car and $\circ$ a vacancy. In
this model cars can move with either velocity $1$ (the $a$ and $s$
processes) or with velocity $2$ (the $u$ process).
%In this model
%at each time step a car can move with velocity $1$ or $2$ with
%probability $a$ and $u$, respectively, provided the gap to the car
%ahead is of length $2$ or more. If that gap is of length $1$, the
%car moves with probability $s$.
The fact that cars perform two types of move will be shown below
to lead to both diffusion and chipping. As in other traffic
models, the general features of the model are revealed only when
the maximal velocity is larger than one.

In what follows we elaborate on the correspondence between this
model and the asymmetric CM. Let us first define the CM more
precisely, and summarize its main features. The CM is defined on a
periodic lattice of size $M$, occupied by $N=\phi M$ particles.
The dynamics is defined through the rates by which two nearest
neighbor sites containing $k$ and $m$ particles, respectively,
exchange particles:
\begin{subequations}
\begin{gather}
\label{eq:cmrates1} (k,m)\mathop{\longrightarrow}^\alpha(k+m,0)\\
\label{eq:cmrates2}
(k,m)\mathop{\longrightarrow}^{q\omega}(k+1,m-1)\qquad
(k,m)\mathop{\longrightarrow}^{(1-q)\omega}(k-1,m+1)\,,
\end{gather}
\end{subequations}
where $q$ controls the bias in the chipping process. Here, for
simplicity, we consider fully left biased diffusion process, which
as will be seen, is the relevant case for the traffic model under
consideration. However, the results quoted below remain valid in
the more general case where diffusion to the right is allowed as
well. It has been shown \cite{Majumdar98A,Majumdar98B,Majumdar98C}
that if the chipping process is symmetric ($q=\half$) there is a
condensation transition at a critical occupancy $\phi_c$, above
which one site becomes macroscopically occupied. Furthermore,
numerical simulations and mean-field studies show that the
occupation probability $p_k$ has the asymptotic form $p_k\sim
z^k/k^\tau$ for large $k$, with $\tau=5/2$. The parameter $z \leq
1$ is determined by the average particles occupancy and serves as
the fugacity. The condensation transition is a result of the fact
that $\tau>2$, for which the distribution $p_k$ cannot sustain
high densities even at $z=1$. This transition is analogous to the
Bose-Einstein condensation. In contrast, if the chipping is
asymmetric ($q\ne\half$) there exists no phase transition at any
occupancy \cite{Rajesh02}. In this case numerical studies indicate
that the domain size distribution has the same form as above, but
here $\tau = 2$. This distribution remains valid at any occupancy,
with $z$ approaching~$1$ as the average occupation is increased.
Thus no condensation transition takes place.

\begin{figure}[t]
\centerline{\epsfig{file=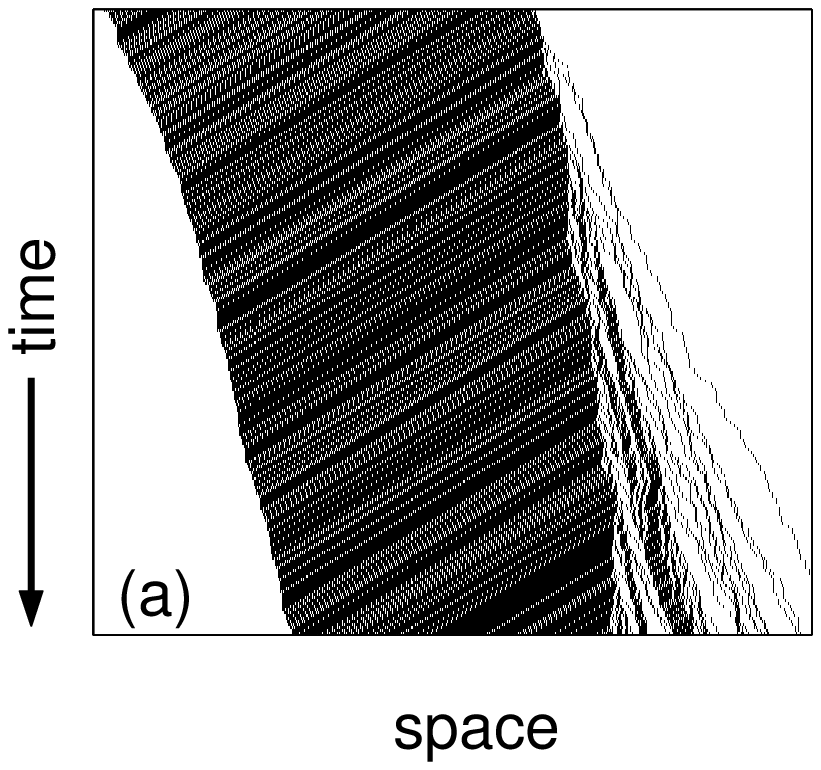,height=6truecm}
\epsfig{file=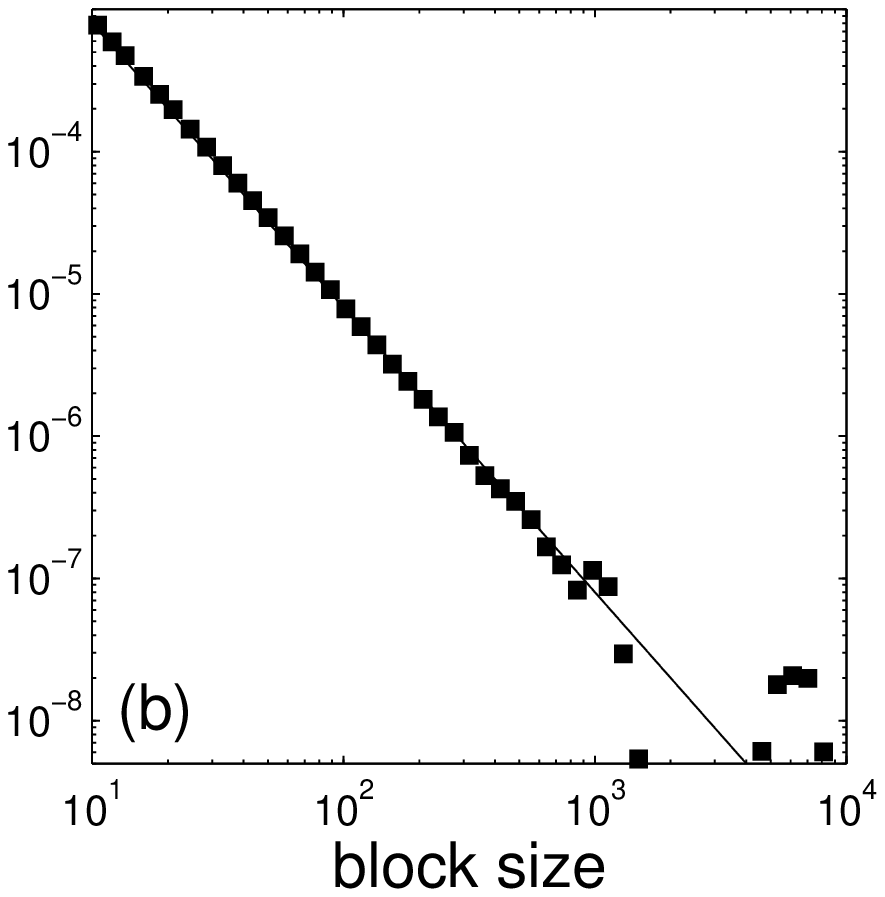,height=6truecm}} \caption{(a) Space-time
configurations of model \eqref{eq:smrates}, with $a=0.5$ and
$u=0.1$. Cars are in black, vacancies in white. (b) Domain size
distribution of the traffic model with $\rho=0.5,a=0.5$ and
$u=0.1$. The solid line has a slope $-2$. Simulations were
performed on a system of size $L=20,000$ and averaged over $4
\times 10^8$ sweeps. \label{fig:sm}}
\end{figure}

Consider now the traffic model \eqref{eq:smrates} in the
deterministic limit $s=1$.  Within this limit it is
straightforward to define a domain (or a jam) as a sequence of
cars and isolated vacancies. These vacancies move
deterministically to the left with velocity $1$. The evolution of
a domain can be described by two processes: (a) a chipping
process, in which a car leaves the domain from its right end with
rate $u$. Such a car leaves two vacant sites behind and is thus
chipped off the domain; and (b) a diffusion process, in which a
vacancy penetrates the domain from its right (with rate $a$) and
advances deterministically to its left, thus shifting its center
of mass one site to the right. In Fig.~\ref{fig:sm}(a) a
space-time configuration of the model at $a=0.5, u=0.1$ is given,
focusing on a single domain. One readily observes the evolution of
the domain through chipping and diffusion.

To test this picture we performed Monte-Carlo simulations of the
traffic model and measured the domain size distribution $p_k$
(Fig.~\ref{fig:sm}(b)). We find that the asymptotic form of $p_k$
is consistent with $k^{-\tau}$ with $\tau=2$, as expected from the
asymmetric CM. The existence of a macroscopic domain, suggested by
the peak at large domains in Fig.~\ref{fig:sm}(b), is thus merely
a finite-size effect.

For $s<1$, vacancies move inside the domain in a non-deterministic
way. They can thus aggregate within a domain, introducing an
additional process in the domain dynamics, namely breaking up of a
domain into two or more domains of comparable size. This process
clearly further reduce the probability of creating a macroscopic
domain. It is thus concluded that the traffic model is not
expected to exhibit a jamming transition at any rate $s$.

\begin{figure}[tb]
\centerline{\epsfig{file=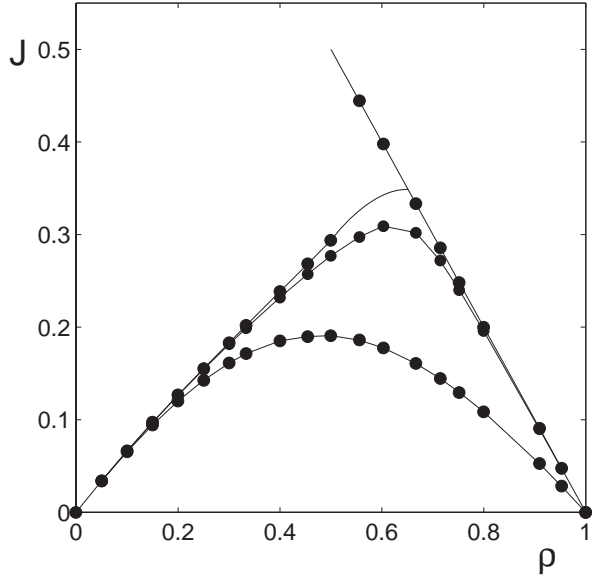,width=8truecm}} \caption{The
fundamental diagram of the extended model, for $a=0.5, u=0.1$ and
$s=0.6, 0.99,$ and $1$ from bottom to top. Simulations are carried
out with $N=100$ particles. Lines serve as a guide to the eye. }
\label{fig:fdem}
\end{figure}

The fundamental diagram of this model as obtained by numerical
simulations is given in Fig.~\ref{fig:fdem}. The diagram exhibits
similar features as that of the simple model of the previous
Section. Here, however, the role of cars and vacancies is
reversed. In the deterministic limit $s=1$  the absorbing state is
found in the high density region, where vacancies move
deterministically in a jam which spans the entire system. At low
densities vacancies can aggregate, and the absorbing state is not
reached. This is the region which corresponds to the asymmetric
CM. In the intermediate region both states can coexist in the
thermodynamic limit, as in the simple traffic model. Due to the
reverse roles played by cars and vacancies, the fundamental
diagram has a lambda-shape rather than the usual inverted-lambda.
For $s<1$ the current is a smooth function of the car density, and
no transition is observed.

\section{Other traffic Cellular Automata}

It would be interesting to examine other traffic cellular automata
within the framework of the chipping model. A particularly simple
and instructive traffic model has recently been introduced in
\cite{Gray01}. Using the same notation as above, the dynamics of
this model is defined by the transition probabilities
\begin{align}
\hspace{-.2cm}
\bullet\bullet\circ\circ&\mathop{\longrightarrow}^\alpha\bullet\circ\bullet\circ
&
\circ\bullet\circ\bullet&\mathop{\longrightarrow}^\beta\circ\circ\bullet\bullet
&
\bullet\bullet\circ\bullet&\mathop{\longrightarrow}^\gamma\bullet\circ\bullet\bullet
&
\circ\bullet\circ\circ&\mathop{\longrightarrow}^\delta\circ\circ\bullet\circ
\,.
\end{align}
The characteristics of traffic flow of this model vary
considerably with the four dynamical rates of the model. In
different limits it exhibits wide moving jams, synchronized flow,
convoys and other features which can be identified in real-life
traffic. In the following we briefly comment on some regions of
this phase space.

The case $\alpha=\gamma=r,\;\beta=\delta=q$ corresponds to the
simplified model introduced in Section~2. Moreover, the case
$\beta=\delta=1$ is qualitatively similar to the case $q=1$ even
when $\alpha \neq \gamma$ \cite{Gray01}. Thus, in this region of
the parameter space the coarse-grained dynamics of the model is
characterized by chipping with no diffusion.

Another interesting case is $\delta=1$, which is the
cruise-control limit of the model. It has recently been argued
that in this case the coarse-grained dynamics is described by the
CM \cite{Levine03}. This correspondence is made by identifying
free-flow domains in typical configurations of the model, whose
dynamics is characterized by both chipping and diffusion
processes. It has also been verified numerically that the gap-size
distribution for large gaps scales as $k^{-2}$, as expected from
the CM. With $\delta<1$ domains can break in their bulk. One thus
expects a stronger suppression of large domains.

Note that if instead of $\delta=1$ one considers $\gamma=1$, the
role played by cars and vacancies is interchanged. In this case
domains correspond to jams, within which deterministic vacancies
are embedded. In this case the distribution of these domains (or
jams) behaves as $k^{-2}$, and thus no macroscopic jam is
expected.

A very interesting region in the parameter space of this model is
the case $\gamma=\delta=1$, corresponding to the cruise-control
limit which is symmetric under car-vacancy exchange (SCC). Here
the model exhibits a low-density absorbing state at $\rho<1/3$ and
a high-density absorbing state at $\rho>2/3$. It has been
suggested \cite{Gray01} that the system exhibits a macroscopic jam
at intermediate densities ($1/3 < \rho < 2/3$) for a particular
region of the $\alpha,\beta$-plane, indicating a jamming phase
transition at some density. Due to the fact that both types of
domains, the high-density and the free-flow ones, are
deterministic, it is possible that this case lies beyond the class
of traffic models described by the CM picture. It would be of
interest to probe this region, as well as other regions of the
parameter space, and study them in the context of the CM picture.

\section{Summary}
\label{sec:summary}

The recently suggested correspondence between traffic dynamics and
the chipping model is reviewed and closely examined in two cases.
The first exhibits chipping but no diffusion. This model is mapped
onto a zero range process (ZRP), which is exactly soluble. The
second is a more general model, whose coarse-grained domain
dynamics exhibit both chipping and diffusion. In both cases it is
shown that when all dynamical rates are non-deterministic the
models do not exhibit a jamming phase transition, but rather a
smooth crossover into the jammed state.

The approach reviewed in this paper could provide a useful tool
for analyzing the behavior of a broader class of traffic models.
Starting from a particular traffic model one should first identify
the {\em domains} which characterize the flow. A domain can either
be a low density segment, termed {\em a gap} or {\em a hole} in
some studies~; a high density segment, termed {\em a jam}~; or a
segment of some other characteristics, defined ad-hoc. A domain of
size $k$ is then associated with a site of the CM occupied by $k$
particles. One then proceeds by examining the evolution of the
domains, and identifying their dynamical processes. In many
traffic models with some deterministic rate these processes are
the diffusion and the chipping processes of the asymmetric CM,
implying that jamming transition does not take place. Usually when
all dynamical rates are not deterministic other dynamical
processes become possible, in which domains can break up into two
or more pieces of comparable sizes. Such processes clearly
disfavor the formation of macroscopic domains. We thus conclude
that jamming transitions are not expected in these models.

A particularly instructive model is the one introduced in
\cite{Gray01}. The correspondence to the chipping model has been
demonstrated in some regions of its parameter space. It would be
interesting to examine more closely other regions of the parameter
space of this model, as well as other traffic models, in more
details.

A different mechanism for condensation transition has been
suggested to take place in some one-dimensional driven systems
\cite{Evans00,Kafri02,Kafri03}. Here the dynamical process
involves only chipping without diffusion. However, in these ZRP
type processes, the chipping rates must decay slowly enough with
the occupation of the departure site for condensation to take
place. Condensation is obtained even when all dynamical processes
are noisy. It would be of interest to examine the relevance of
this mechanism to traffic models. This mechanism could lead to a
jamming transition, which is absent in the asymmetric chipping
model.

\ack We thank M.R. Evans and J.L. Lebowitz for useful discussions.
The support of the Israel Science Foundation (ISF) and NSF Grant
DMR 01-279-26 is gratefully acknowledged.

\end{document}